\documentclass[10pt,a4paper]{article}

\usepackage{times}

\usepackage{epsfig,wrapfig}
\usepackage{epstopdf} 

\usepackage{fancyhdr}

\usepackage[lmargin=2.5cm, rmargin=2.5cm,tmargin=3.50cm,bmargin=3.50cm]{geometry}
\usepackage{indentfirst}
\usepackage{subcaption}

\usepackage{titlesec}
\titleformat{\section}[hang]
  {\centering}{\thesection}{1ex}{\normalsize \textsc}
\titleformat{\subsection}[hang]
  {}{\thesubsection}{1ex}{\normalsize \textit}

\renewcommand{\thesection}{ \normalsize \textnormal{\Roman{section}.}}
\renewcommand{\thesubsection}{\normalsize \textnormal{\textsc{\textit{\Alph{subsection}.}}}}

\def\e{\begin{equation}}
\def\f{\end{equation}}
\def\_#1{{\bf #1}}

\def\.{\cdot}

\begin{document}

\title{\large \textbf{Metamirrors}}
%
\def\affil#1{\begin{itemize} \item[] #1 \end{itemize}}
\author{\normalsize \bfseries \underline{V.~S.~Asadchy}$^{1,2}$, Y.~Ra'di$^1$ and S.~A.~Tretyakov$^1$
}
\date{}
\maketitle
\thispagestyle{fancy} 
\vspace{-6ex}
\affil{\begin{center}\normalsize $^1$Department of Radio Science and Engineering, Aalto University\\ P.O.~13000, FI-00076 Aalto, Finland\\
$^2$Department of General Physics, Gomel State University, 246019, Belarus\\
viktar.asadchy@aalto.fi
 \end{center}}

\begin{abstract}
\noindent \normalsize
\textbf{\textit{Abstract} \ \ -- \ \
We introduce the concept  of non-uniform metamirrors
(full-reflection metasurfaces) providing full control of reflected
wave fronts independently from the two sides of the mirror.
Metamirror is a single planar array of electrically small
bianisotropic inclusions. The electric and magnetic responses of the
inclusions enable creating controlled gradient of phase
discontinuities over the surface. Furthermore, presence of
electromagnetic coupling in the inclusions allows independent
control of reflection phase from the opposite sides of the mirror.
Based on the proposed concept, we design and simulate metamirrors
for highly efficient light bending and near-diffraction-limit
focusing with a sub-wavelength focal distance.}
\end{abstract}

\section{Introduction}

Metasurfaces are electrically thin composite layers with engineered
electromagnetic properties. Through the use of metasurfaces, full
control over amplitude and phase of reflection and transmission
becomes possible. Transparent metasurfaces \cite{transparent},
absorbing sheets \cite{absorption}, full-reflection layers with
arbitrary phase of reflection (metamirrors, considered here)
\cite{metamirrors} are only some examples demonstrating novel and
exciting properties unavailable with natural materials. Using
in-plane non-uniform metasurfaces it becomes possible to emulate the
properties of transmit-arrays \cite{snell}-\cite{grady}. Another
capability provided by non-uniform metasurfaces is
near-diffraction-limit focusing \cite{shalaev,alu}. As it was shown
in \cite{metamirrors} and \cite{alu}, in order to create a phase
variation of the \mbox{reflection/transmission} spanning a 2$\pi$
range, both tangential electric and magnetic currents on the
metasurface are needed. Thus, such metasurfaces must possess a
finite thickness to provide necessary magnetic loop currents.
Metasurfaces supporting only electric surface currents
\cite{snell,grady,shalaev} have very low efficiency (less than
25$\%$). Furthermore, known metasurfaces with only electric surface
currents operate only with one specific polarization of incidence.
Recent works \cite{grbic} and \cite{alu} have demonstrated highly
efficient transmit-arrays of polarizable particles providing both
electric and magnetic polarization currents to generate prescribed
wave fronts. However, they operate only with one polarization of
incidence and allow control only over transmission.

Here, we present a new metasurface concept  (so-called metamirror)
providing full control of reflected wave fronts. Conventional
sub-wavelength reflectarrays \cite{Pozar} consist of closely spaced
electrically small patches on a grounded substrate. But the ground
plane forbids transmission at all frequencies and limits properties
from the opposite side. We utilize the idea of uniform
full-reflection sub-wavelength metasurfaces proposed in
\cite{metamirrors} and develop it for realization of specific
non-uniform phase distributions of reflected plane waves. The design
represents a single planar array of specifically shaped resonant
bianisotropic particles possessing omega electromagnetic coupling.
The electromagnetic coupling is crucial in order to create all
necessary phase variations of the reflection from the two sides of
the metamirror. It allows full tailoring of co-polarized reflection
for  arbitrary polarized normally incident plane waves. At the same
time, due to precise tuning, the structure can approach 100$\%$
efficiency. The designed metasurfaces are electrically thin since
the inclusions have dipolar response. Moreover, the omega
electromagnetic coupling enables engineering asymmetric response
from metasurfaces (e.g., negative refraction from one side and
sub-wavelength focusing from the opposite side of the surface)
\cite{metamirrors}. Although here we realize the concept of
functional metamirrors only in the microwave frequency range, it can
be applied to higher frequencies as well.

\section{Anomalous reflection}

We design a metasurface to  efficiently reflect normally incident
plane waves to an arbitrarily chosen angle $\phi=45^{\circ}$ from
the normal. The metasurface consists of a sub-wavelength periodical
planar array of bianisotropic omega inclusions.  Since the array
period is sub-wavelength, the induced moments can be modelled as
surface-averaged electric and magnetic surface currents that radiate
secondary plane waves. At the resonance, the metasurface with
adjusted inclusion density totally reflects incident waves. Precise
tuning the phases of scattered waves from each inclusion allows us
to tailor the front of the reflection from the metasurface.

\begin{figure}[h!]
\centering
\begin{subfigure}{0.45\columnwidth}
  \centering
  \includegraphics[width=\columnwidth]{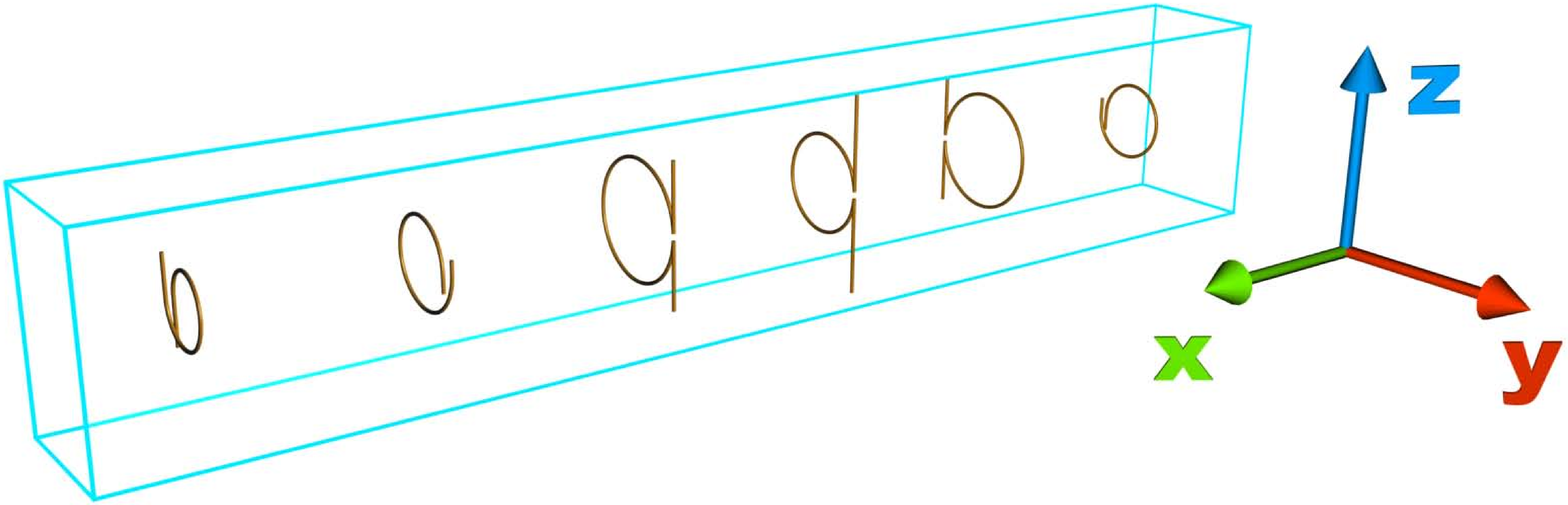}
  \caption{}
  \label{fig:cell}
\end{subfigure}\qquad
\begin{subfigure}{0.36\columnwidth}
  \centering
  \includegraphics[width=\columnwidth]{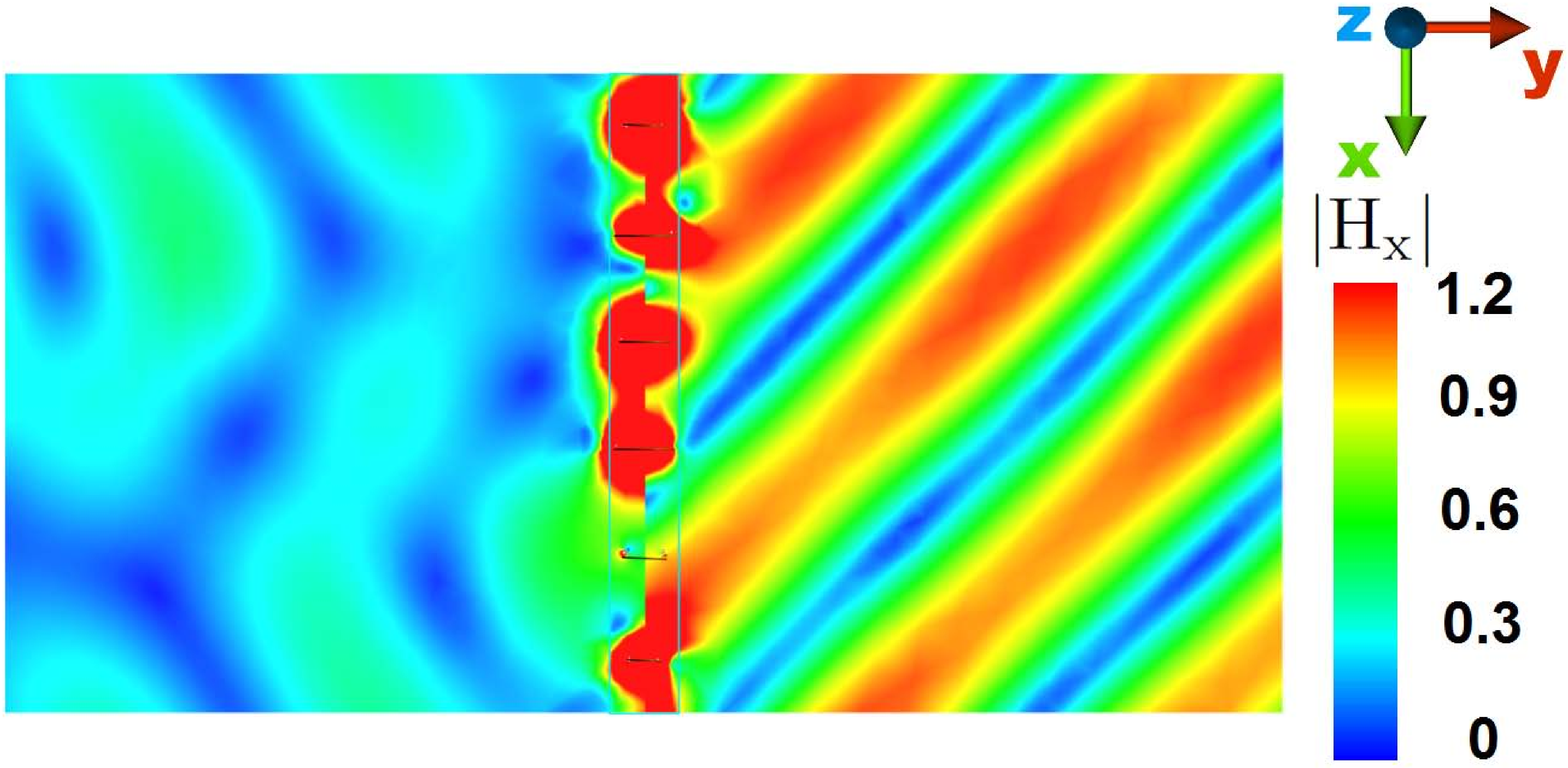}
  \caption{}
  \label{fig:deflection}
\end{subfigure}
\caption{(a) A unit cell of the metasurface providing a linear phase variation  of the reflection spanning a 2$\pi$ range. The dielectric substrate is not shown for clarity. (b) Magnetic field distribution (normalized to the magnetic field of the incident wave) of the transmitted (on the left side) and reflected (on the right side) waves. The metasurface is illuminated by a normally incident $z$-polarized plane wave at 1.15~GHz.}
\label{fig:test}
\end{figure}

In order to realize  anomalous reflection, a linear phase variation
along the interface is required. Using the principle of phased
arrays, it is simple to show that the reflected wave front is
deflected to an angle $\phi$ if the metasurface provides a linear
phase variation spanning 2$\pi$ range with the periodicity
$d=\lambda/\sin\phi$. The use of sub-wavelength inclusions ensures
homogeneous (in the sense that the averaged electromagnetic response
varies smoothly over the structure) phase variation along the
interface.
In our design for the microwave frequency range we use copper wire
omega  particles embedded in a dielectric material with the
permittivity $\epsilon_{\rm r}=1.03$ and thickness 40~mm along the
$y$-axis (see Fig.~\ref{fig:cell}).
The super-unit-cell  (shown as a blue box) consists of 6 particles
providing discrete phase shifts that approximate homogeneous
electromagnetic response. The operating frequency of 1.15~GHz is
chosen. At this frequency the cell period along the $x$-axis is
$d=\lambda/\sin45^{\circ}=369$~mm. The unit cell is repeated also
along the $z$-axis with the periodicity $d/6=61.5$~mm. The shape and
dimensions of the particles were optimized by numerical simulations
in order to create the required discrete phase shifts: $0^{\circ},
60^{\circ}, 120^{\circ}, 180^{\circ}, 240^{\circ}, 300^{\circ}$. An
incident plane wave impinges on the metasurface from the
$+y$-direction with the electric field parallel to the $z$-axis.
Fig.~\ref{fig:deflection} shows magnetic field distribution of the
reflected and transmitted waves (Ansoft HFSS simulations). The
 phase front of the reflection is indeed planar
and deflected $45^{\circ}$ with respect to the normal. 10.3$\%$ of
the incident power is transmitted and 4.4$\%$ is absorbed by copper
wires. Reflectance in this case is 85.3$\%$ and can be further
increased by inclusions optimization. Despite the thickness  of the
designed metasurface $\lambda/7.6$ is already quite small, it can be
decreased  further by choosing different particle shapes.

\section{Near-diffraction-limit focusing at sub-wavelength distances}

Another example  demonstrating the flexibility and uniqueness of
non-uniform metamirrors is focusing metasurfaces  showing extremely
strong wave-gathering ability. The metamirror concept allows us to
design a conceptually new kind of a lens: one consisting of an
ultimately thin single layer of resonant dipolar inclusions
providing near-diffraction-limit focusing of electromagnetic energy
at any focal distance and at any point. Moreover, asymmetric
response from different sides of the lens can be achieved providing
the extreme freedom for engineering. Since the metalens is planar,
it does not possess spherical aberrations.  The proposed concept can
be applied for any frequency range.

\begin{figure}[h!]
\centering
\begin{subfigure}{0.28\columnwidth}
  \centering
  \includegraphics[width=\columnwidth]{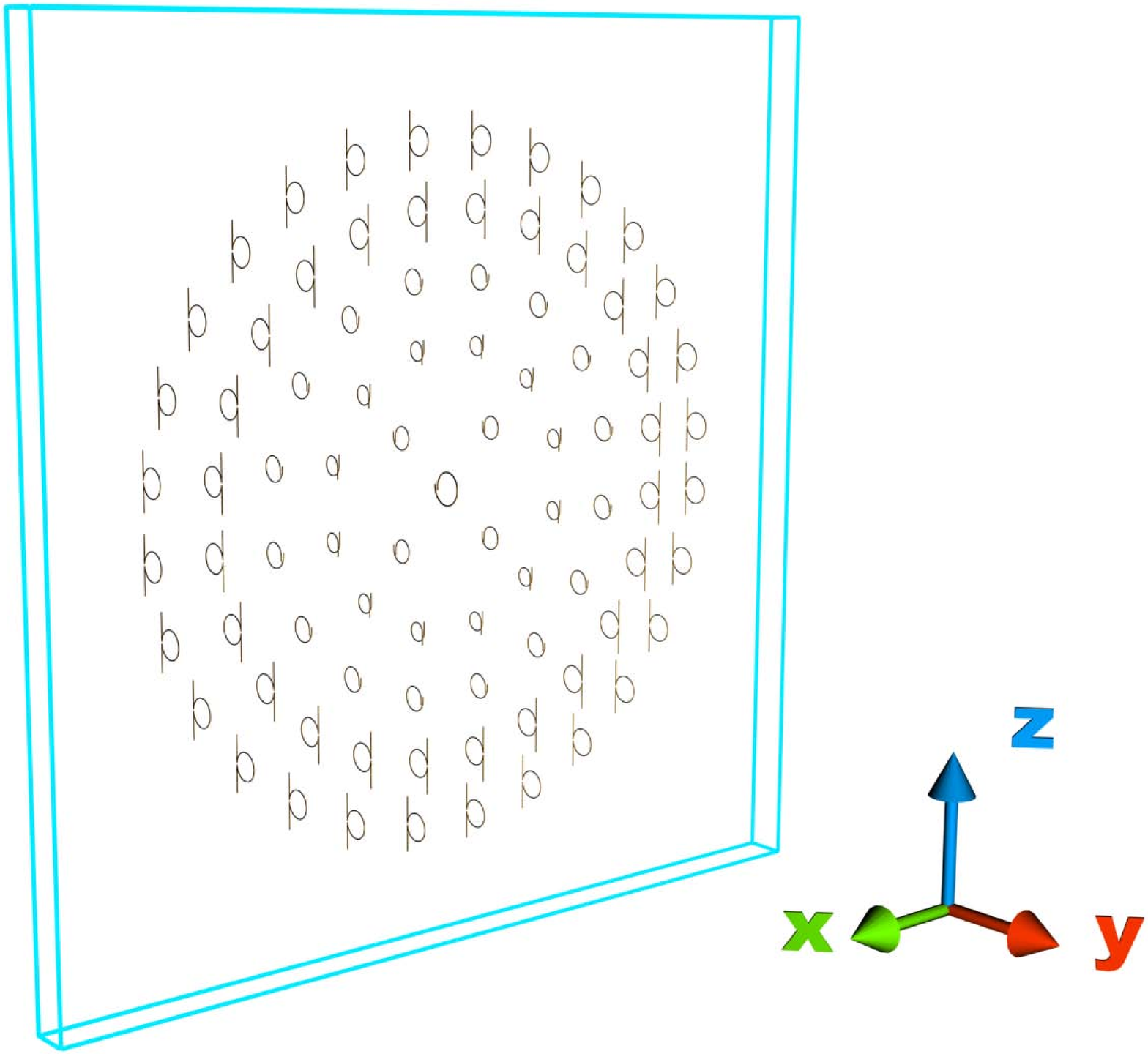}
  \caption{}
  \label{fig:lens}
\end{subfigure}\qquad\qquad
\begin{subfigure}{0.35\columnwidth}
  \centering
  \includegraphics[width=\columnwidth]{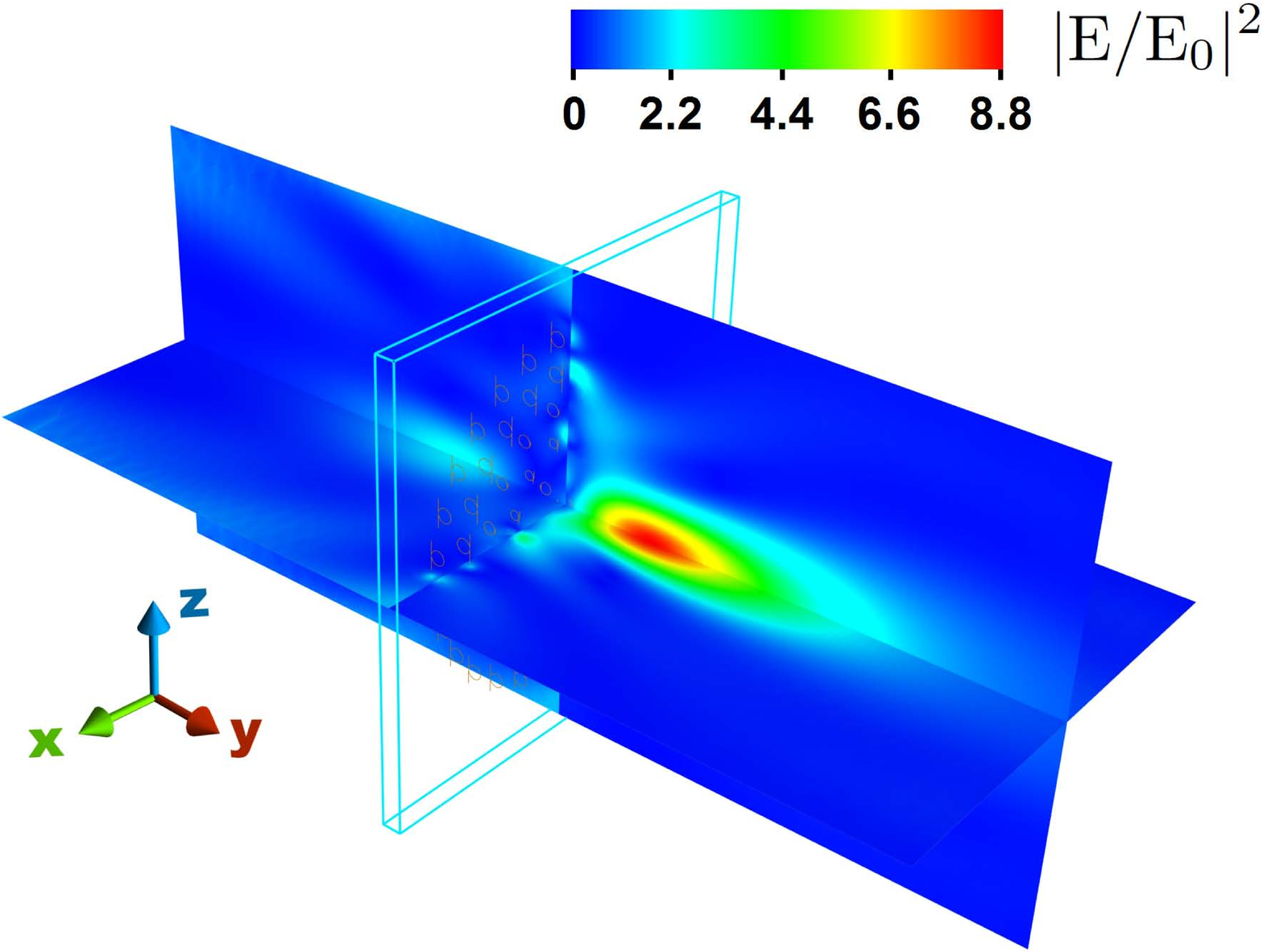}
  \caption{}
  \label{fig:focusing}
\end{subfigure}
\caption{(a) A single-layer metalens. The blue box denotes the dielectric substrate. (b) Power density distribution (normalized to the incident power density) of the reflection (the $+y$ half-space) and transmission (the $-y$ half-space). Power intensity maps are depicted on two orthogonal cross-section planes (the $xy$ and $yz$-planes).}
\label{fig:test2}
\end{figure}

Realization of lens response requires certain phase  variations
along the surface  ensuring that the scattered fields from all
inclusions constructively interfere in the desired point. To
demonstrate the lensing effect, we designed a single-layer
metamirror composed of 6 concentric arrays of bianisotropic omega
inclusions (see Fig.~\ref{fig:lens}).
The concentric arrays provide the desired  phase shifts from the
center to the edges: $-80^{\circ}, -54^{\circ}, -2^{\circ},
65^{\circ}, 140^{\circ}, -140^{\circ}$. The working frequency and
the supporting dielectric slab (marked as a blue box on the figure) are the same as in the previous
example. The dimensions of the metalens: the radius is $1.4\lambda$ and the thickness is $\lambda/7.6$. In order to demonstrate the power of the concept, the inclusions are designed to confine electromagnetic energy in a spot
at an extremely small focal distance $0.6\lambda$. An incident plane
wave impinges on the metalens from the $+y$-direction (the electric
field parallel to the $z$-axis). Fig.~\ref{fig:focusing} shows the
simulated power distribution of the reflection and transmission from
the metalens at 1.155 GHz. Very weak transmission in the far-zone is
caused by diffraction effects on the edges of the metasurface. The
metalens effectively reflects the wave and focuses it tightly near
the diffraction limit to a spot of only $2.8\lambda\times0.9\lambda$
size ($1/e^2$ beamwidth). The extremely strong focusing ability of
the designed metalens provides the focal length of only
$0.73\lambda$ and high energy gain of 8.8 in the spot. The f-number
(the ratio of the lens's focal length to its diameter) for the
designed metalens comes to 0.26 and can be further decreased by
increasing the lens' diameter. Such small f-number of the metalens
allows of gathering more power and generally providing a brighter
image. These parameters of the designed lens significantly exceed
those of other metamaterials-based lenses, e.g. \cite{shalaev,alu}.

\section{Conclusion}
Our results show that  the metamirror concept enables extreme
independent control over reflection from the two sides of the
surface and allows of designing ultimately thin, 100$\%$-efficient,
polarization-insensitive devices with desired properties. Similar
full control over transmission can be accomplished through the use
of the Babinet-inverted design.



\end{document}